\documentclass[preprint,aps,tightenlines]{revtex4}

\usepackage{epsfig}
\begin{document}
\title{A Relationship Between Parametric Resonance and Chaos}
\author{R. Kobes}
\email{randy@theory.uwinnipeg.ca}
\altaffiliation{ Winnipeg Institute for
Theoretical Physics, Winnipeg, Manitoba, R3B 2E9 Canada}
\affiliation{Department of Physics, University of Winnipeg, Winnipeg, 
Manitoba, R3B 2E9 Canada}
\author{S. Pele\v{s}}
\email{peles@theory.uwinnipeg.ca}
\altaffiliation{ Winnipeg Institute for
Theoretical Physics, Winnipeg, Manitoba, R3B 2E9 Canada}
\affiliation{Department of Physics, University of Manitoba, Winnipeg, 
Manitoba, R2T 2N2 Canada}

\begin{abstract}
In this paper we study two types of exponential instability -- parametric
resonance and chaos. We show that a given equation may produce
chaos or parametric resonance, depending how the problem is defined.
In so doing we establish a relationship between the 
Floquet indices (associated with parametric resonance) 
and Lyapunov exponents (associated with chaos).
\end{abstract}

\maketitle

\section{Introduction}

Parametric resonance is a phenomenon that occurs in various cosmological and
high energy physics models. It manifests itself
as a rapid growth of a physical field at an
exponential rate. Recently this phenomenon has been
used to explain some physical processes such as reheating in the 
early universe \cite{kls,kaiser1} and in phase transitions 
in disordered chiral condensates \cite{kaiser2}.
At the same time a lot of attention has been given to the
study of chaotic systems, i.e. systems whose trajectories in phase space
diverge
exponentially, but at the same time remain within a bounded region. 
As both types of systems are described by an exponential type of
instability one might expect a relationship between the two,
and in this paper we investigate quantitatively just such a
relationship. We show that for a
system exhibiting parametric resonance it is possible to construct an
equivalent chaotic system, although the converse is not guaranteed.

\subsection{Floquet Index}

{}From the general theory of differential equations we know that any
second-order
linear differential equation
\begin{equation}
        \frac{d^2y}{dt^2}+f(t)\frac{dy}{dt}+g(t)y=0,
\label{eq:flocquet}
\end{equation}
will have two linearly independent solutions. According to Floquet's theorem
\cite{mb}
if $f(t)$ and $g(t)$ are functions periodic in $t$ with a period $T$, then
those solutions will have form:
\begin{equation}
        y(t)=e^{\mu t}P(t),
\label{eq:solution}
\end{equation}
where $P(t)$ is periodic function with period $T$, as well.
Therefore, stability of the solution (\ref{eq:solution}) is entirely determined
by the
exponent $\mu$, which is also called Floquet exponent or Floquet index.
There is no general procedure for estimating Floquet exponent, however there
are
a lot of particular cases such as the Mathieu equation where an
extensive analysis of the Floquet indices has been done.

\subsection{Lyapunov Exponents}

Lyapunov exponents are a quantitative indication of chaos, 
and are used to measure the
average rate at which initially close trajectories in phase space diverge from
each other. The Lyapunov exponent is usually defined as:
\begin{equation}
        \lambda_{i}=\lim_{t \rightarrow \infty}\frac{1}{t}
        \ln \frac{\epsilon_{i}(t)}{\epsilon_{i}(0)}
\label{eq:lyapun}
\end{equation}
where $\epsilon_{i}(t)$ denotes the 
separation of two trajectories, and the index $i$
denotes the direction of growth in phase space or 
Lyapunov directions \cite{Wol84}.
If at least one Lyapunov exponent is
positive, then
trajectories of the system will separate at an exponential rate and we
say that system is chaotic. This will manifest itself
in a high sensitivity to a
change in the initial conditions.
Since the motion of chaotic systems is performed within a
bounded region in phase
space, besides an exponential stretch 
some kind of fold of the trajectory must occur.
Because of that, the Lyapunov directions $i$ in phase space
are not constant in time. They tend to rotate, and at the same time to
collapse towards the  direction of the most rapid growth.
A powerful and very general algorithm for determining Lyapunov exponents
is presented in Ref.~\cite{Ben80,Shi79}, and its implementation, 
which we use for our
calculations, in Ref.~\cite{Wol84}.

The algorithm is based on an
analysis of the evolution of an orthonormal basis in phase space
along a trajectory. In the
linear approximation directions of growth in phase
space
will remain orthogonal, therefore allowing one to measure the growth
directly.
In Ref. \cite{Wol84} it
is suggested to describe evolution of the orthonormal basis
by linearized equations and use a small enough time step so this approximation
remains valid, and so
we can neglect the effect of the Lyapunov directions collapsing.
After the growth was measured the basis is re-orthonormalized, and the
procedure is repeated for the next small enough time interval. If the
limit of Eq.~(\ref{eq:lyapun})
exists, then the average of the
measured growths should converge to its asymptotic value.
This is a robust numerical algorithm and it works for almost every problem
whose governing equation is known.
It is not suitable though for unbounded systems that diverge exponentially,
since the limits of numerical accuracy
may be exceeded before the average growth starts converging.

\section{Mathieu Equation}

\begin{figure}
\centerline
{
        \epsfig{file=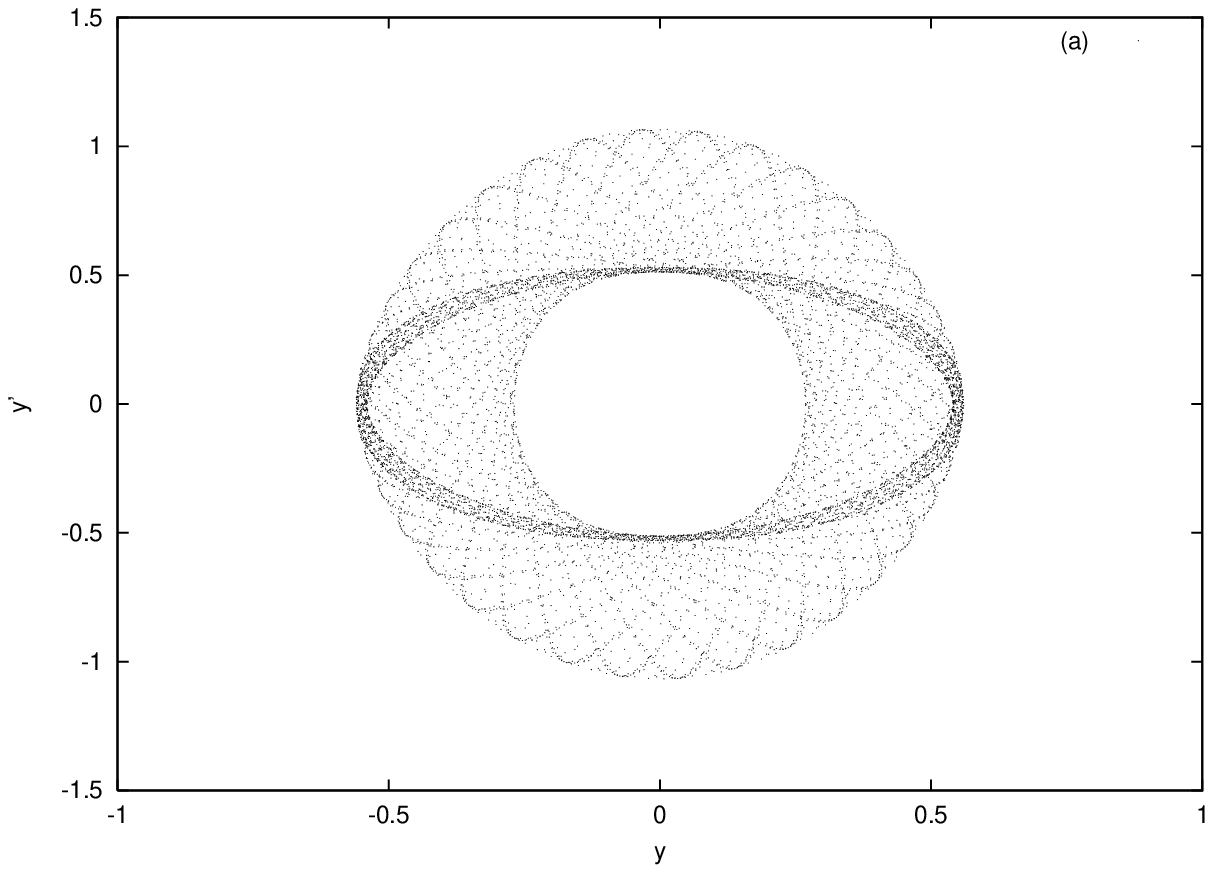, height=4.5cm}
        \epsfig{file=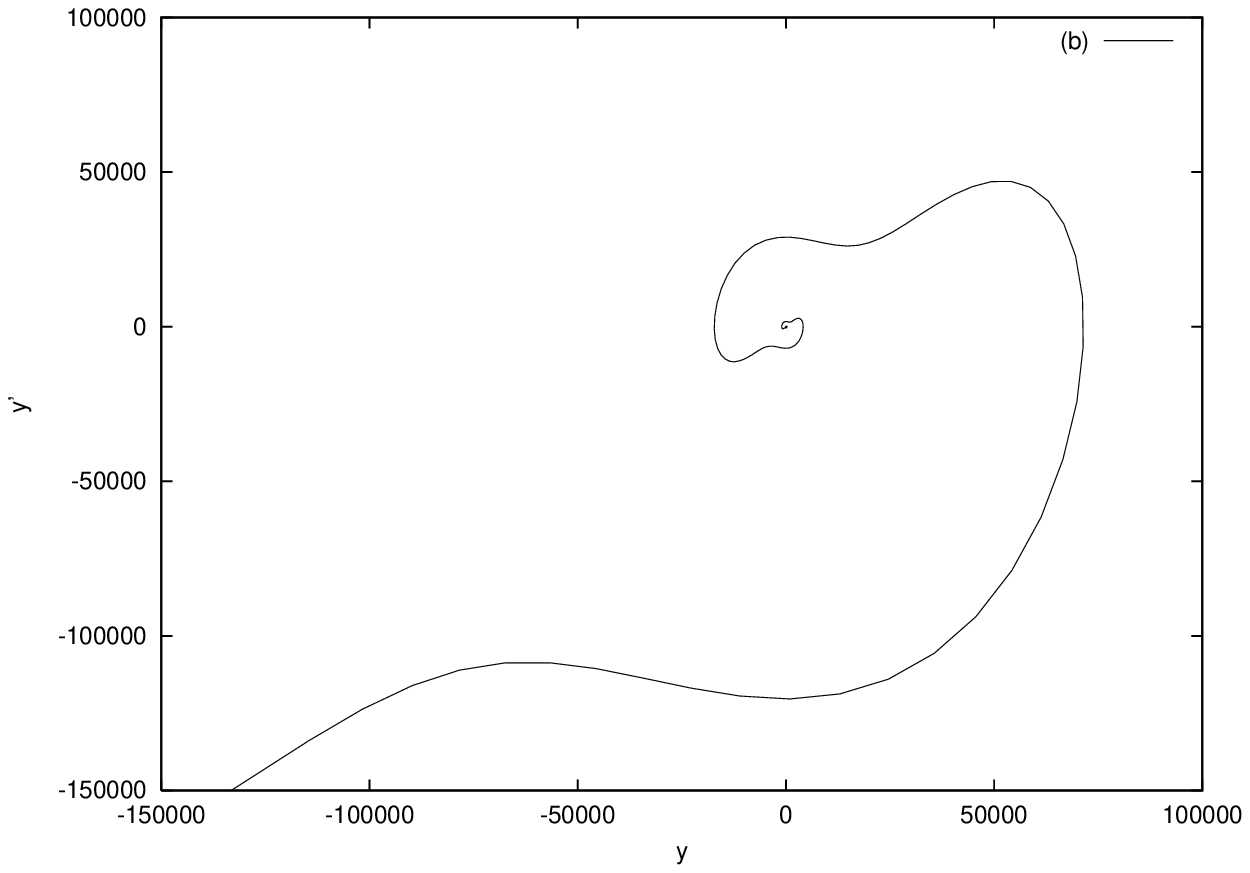, height=4.5cm}
}
        \label{fig:phase}
        \caption{Phase space trajectory for the Mathieu equation.
        (a) $A=2.5$ and $q=1$, (b) $A=1$ and $q=1$.}
\end{figure}

Perhaps the simplest model that exhibits parametric resonance 
is the Mathieu equation (\ref{eq:mathiew}), which was 
originally used to describe small oscillations of a
pendulum with vertically driven base.
\begin{equation}
        y''+(A-2q \cos2t)y=0
\label{eq:mathiew}
\end{equation}
This is a Floquet type equation with two parameters $A$ and $q$, and
it has a solution of the form of Eq.~(\ref{eq:solution}). The
value of the Floquet index
$\mu$ depends on the equation's parameters. For certain values of
$A$ and $q$ (e.g. $A=2.5$ and $q=1$) Floquet exponents would be purely
imaginary, meaning that the solutions of Eq.~(\ref{eq:solution}) 
will both be periodic. Therefore, the
solution will be stable, and its trajectory in phase space will remain
within a bounded region (see Fig.~\ref{fig:phase}a ).
Otherwise, both Floquet exponents will be purely real, and one of them hence
positive. A solution with a positive Floquet exponent is unstable and grows
exponentially (see Fig.~\ref{fig:phase}b). In some physical models 
such growth of the field $y$ can be
interpreted as a massive production of certain particles.
This is also referred to as parametric resonance.

\begin{figure}
\centerline
{
        \epsfig{file=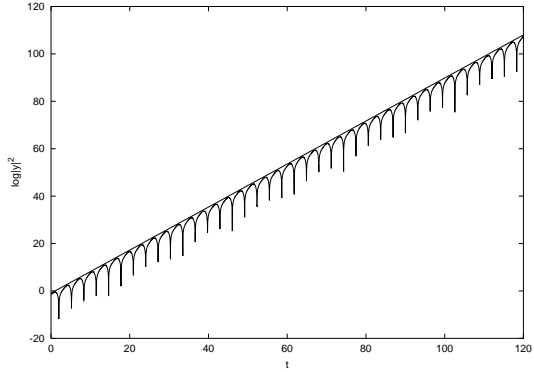, height=5cm}
}
        \label{fig:time}
        \caption{Exponential growth of the solution for the Mathieu equation.
        The Floquet exponent is estimated from the slope to be
$\mu=0.453\pm0.003$}
\end{figure}

The rate of exponential growth of
the solution (i.e. a positive Floquet exponent) can be determined from
the graph for $\log|y|^2$ plotted against time $t$. Since the term in the
solution containing a positive exponent is dominant, the
slope of envelope of
the graph will yield numerical value of $2\mu$. From Fig.~(\ref{fig:time})
it is found that $\mu=0.453\pm0.003$, where the 
parameters are chosen to be $A=1$ and $q=1$.
Regions of stability in parameter space of the Mathieu equation have been
very well studied (see, for example, Ref.~\cite{Abr65}). There are 
 bands of stability and 
instability in the parameter space, and their boundaries are
continuous curves.

\begin{figure}
\centerline
{
        \epsfig{file=periodic.eps, height=4.5cm}
        \epsfig{file=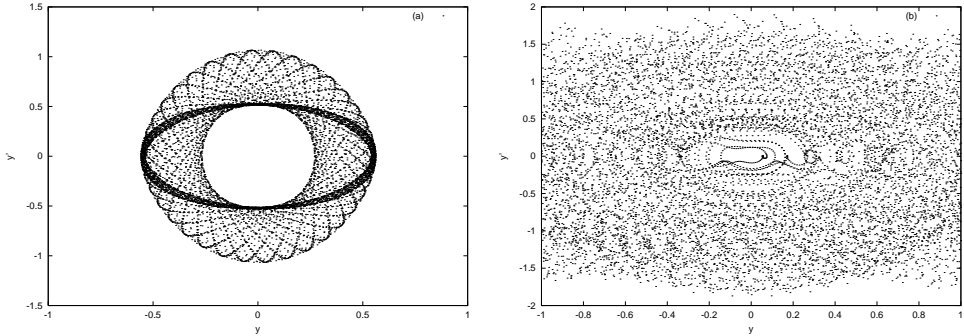, height=4.5cm}
}
        \label{fig:phase2}
        \caption{Phase space trajectory for the
          Mathieu equation with appropriate
        winding condition. The periodic solution (a), $A=2.5$ and $q=1$,
        remains unchanged, while the unstable solution (b), $A=1$ and $q=1$,
        becomes chaotic.}
\end{figure}
In analogy with interpreting $y$ as an angle, we impose suitable ``winding''
conditions on the solution of Eq.~(\ref{eq:mathiew}) so that
it always stays within
segment $[-1,1]$. There is no physical motivation to interpret $y$ as an
angle though, unless it stays within the limits of a small angle approximation.
With this additional restriction imposed, both stable and unstable solutions
are bounded, so parametric resonance does not occur. The stable 
solution remains
periodic and exhibits the same behavior as before (see Fig.~\ref{fig:phase}).
The unstable solution, on the other hand, instead of parametric resonance,
exhibits chaotic-like behavior (see Fig.~\ref{fig:phase2}) which manifests in
high sensitivity in change of initial conditions.
For this solution we estimated the Lyapunov spectrum, and found 
the positive Lyapunov
exponent to be $\lambda_{1}=0.453\pm0.001$, which is the
same as the Floquet exponent.
This result could be anticipated
because the first Lyapunov direction always point to
the direction of the fastest growth in phase space. For the Mathieu equation
this 
growth is entirely described by the solution with a positive Floquet exponent.
The linearization procedure in the
algorithm for the Lyapunov exponents calculation
will in a sense ``unwind'' the trajectory, so that
the exponential divergence
measured by the Lyapunov exponent has to be the same as that described
by the Floquet index.

\section{Parametric Resonance Model}

\begin{figure}
\centerline
{
        \epsfig{file=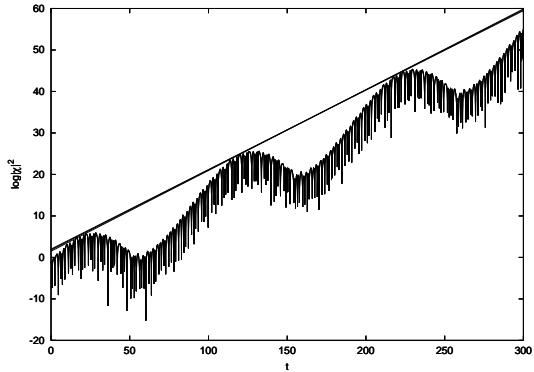, height=5cm}
}
        \label{fig:param}
        \caption{Plot of $\log|\chi|^{2}$ vs. $t$ shows how the 
          Floquet exponent
          changes during the time. The average 
          Floquet exponent is estimated from
        the graph to be $\mu_{\chi}=0.0973\pm0.0007$.}
\end{figure}

If the parameters of the Mathieu equation Eq.~(\ref{eq:mathiew}) 
are not constant,
but rather some functions
of time, then the solution will eventually switch between regions of
stability and instability in parameter space, and therefore phases of
quasiperiodicity and exponential growth will interchange during that time.
Here is a somewhat simplified system of equations which illustrates such
behavior. Eqs.~(\ref{eq:hi}, \ref{eq:fi})
may be used to describe the
decay of $\phi$-particles into $\chi$-particles.
\begin{equation}
        \ddot{\chi}+H\dot\chi+(m^{2}+g\phi^{2})\chi=0
\label{eq:hi}
\end{equation}
\begin{equation}
        \ddot{\phi}+(A-2q\cos2t)\phi=0
\label{eq:fi}
\end{equation}
Eq.~(\ref{eq:hi}) is a Floquet--type of
equation, with parameters $m$ and $g$ set
near the boundary between the stability and instability regions. 
Eq.~(\ref{eq:fi}) is a
Mathieu equation which is coupled to Eq.~(\ref{eq:hi}).
If we set the parameters $A$ and $q$ so that Eq.~(\ref{eq:fi}) has a
periodic solution, then the
term $g \phi^{2}$ will periodically drive Eq.~(\ref{eq:hi})
between its stability and instability regions, and therefore its Floquet
exponent
will change periodically in time. This is shown in Fig.~\ref{fig:param},
where
$\log|\chi|^{2}$ is plotted versus time. Although the
Floquet exponent changes
periodically in time, the system spends more time in a region of
instability, and
parametric resonance occurs. The
average value of the Floquet exponent has a positive real part,
and its numerical
value can be estimated directly from Fig.~\ref{fig:param}.

\begin{figure}
\centerline
{
        \epsfig{file=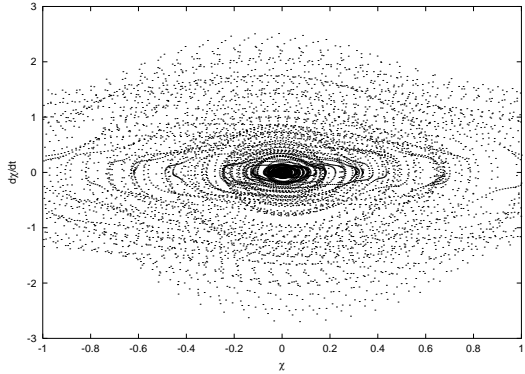, height=5cm}
}
        \label{fig:phase3}
        \caption{ Phase space projection of the chaotic pseudo-attractor of
        Eq.~(\ref{eq:hi}).}
\end{figure}

Now, let us impose the same ``winding'' conditions like that for the Mathieu
equation, so that $\chi \in [-1,1]$. As before, parametric resonance
will not occur, but the field $\chi$ will
exhibit chaotic--like behavior instead. In order to find the 
Lyapunov exponent
spectrum for this system we need to perform our calculation in 5-dimensional
phase space $(\dot \chi, \chi, \dot \phi, \phi,t)$. We found two exponents in
the spectrum to be positive, and their sum to be
$\lambda_{1}+\lambda_{2}=0.0973 \pm 0.0006$, which agrees with the
value for
average Floquet exponent.

Again, this is an expected result, considering the algorithm for 
estimation of the Lyapunov exponents. The sum 
of all positive Lyapunov exponents is an average rate of
exponential divergence of the solution of Eq.~(\ref{eq:hi}). This should
be the same as the rate estimated from Fig.~\ref{fig:param}, as the slope
there is determined by the average value of Floquet exponent.

If we, however, substitute a chaotic solution of the Mathieu equation into
Eq.~(\ref{eq:hi}), the
system will exhibit a very complex behavior. 
The system will chaotically switch
between stability and instability regions so it will be impossible to predict
any kind of resonant behavior due to a high sensitivity to change in initial
conditions. Furthermore, Eq.~(\ref{eq:hi}) will not be a Floquet equation
any more, and its solution will not have the simple form 
of Eq.~(\ref{eq:solution}).

\section{Conclusions}

We demonstrated here that parametric resonance and chaos are two types of
exponential instability which are mutually exclusive but related.
Starting from a model that satisfies Floquet's theorem and is in
a region of parameter space which is exponentially unstable
(with a positive Floquet index), we showed that imposing a
``winding'' type of boundary condition on the field to restrict
it to lie within a certain range leads to a model exhibiting
chaos, and hence with at least one positive Lyapunov exponent.
A quantitative measure of the
exponential divergence rate in the two related models of the cases
we studied shows that the Floquet exponent is equal to the 
Lyapunov exponent. Some extensions and applications of this
correspondence between these two types of instabilities are
currently being studied.

\section*{Acknowledgements}
This work was supported by the Natural Sciences and
Engineering Research Council of Canada.

\end{document}